\documentstyle[aps]{revtex}
\begin{document}
\newcommand{\beq}{\begin{equation}}
\newcommand{\eeq}{\end{equation}}
\newcommand{\beqn}{\begin{eqnarray}}
\newcommand{\eeqn}{\end{eqnarray}}
\newcommand{\bmath}{\begin{mathletters}}
\newcommand{\emath}{\end{mathletters}}
\twocolumn[\hsize\textwidth\columnwidth\hsize\csname @twocolumnfalse\endcsname
\title{Spin Hall effect}
\author{J. E. Hirsch}
\address{Department of Physics, University of California, San Diego\\
La Jolla, CA 92093-0319}
 
\date{February 24, 1999} 
\maketitle 
\begin{abstract} 
It is proposed that when a charge current circulates in a paramagnetic
metal a transverse spin imbalance will be generated, giving rise
to a 'spin Hall voltage'. Similarly, that when a spin current circulates
a transverse charge imbalance will be generated, hence a Hall voltage,
in the absence of charge current and magnetic field. Based on these
principles we propose an experiment to generate and detect a 
spin current in a paramagnetic metal.

\end{abstract}
\pacs{}
\vskip2pc]

Consider the 'spontaneous' or 'anomalous' Hall effect\cite{hurd}.
In ferromagnetic metals, the Hall resistivity (transverse electric
field per unit longitudinal current density) is found to be
empirically fitted by the formula
\beq
\rho_{H} = R_{o}B + 4\pi R_{s}M
\eeq
(in cgs units), with B the applied magnetic field and M the magnetization 
per unit volume.  
$R_{o}$ is the ordinary Hall coefficient and $R_{s}$ the 'anomalous'   Hall
coefficient, experimentally found to be generally substantially larger than
the ordinary Hall coefficient as well as strongly temperature-dependent.
Within models that assume that the electrons giving rise to magnetism
in ferromagnetic metals are itinerant, a variety of mechanisms have been
proposed to explain the origin of the coefficient $R_s$. These
include skew scattering by impurities and
phonons, and the 'side jump' mechanism. In early
work it was also proposed that the effect will  arise in the
absence of periodicity-breaking perturbations\cite{karp}, but this is
generally believed not to be correct\cite{hurd}.

In this paper we will not discuss the origin of the anomalous Hall
effect\cite{hirs}. Rather, we take the existence of the effect in 
ferromagnetic metals as experimental proof that electrons carrying
a spin and associated magnetic moment experience a transverse force
when they are moving in a longitudinal electric field, for any of the
reasons listed above or others. If there is a net magnetization in the
system there will be a magnetization current associated with the flow
of electric current, and the transverse force will give rise to a 
charge imbalance in direction perpendicular to the current flow and
hence to an anomalous Hall effect.

Consider then the situation where no magnetization exists, that is,
a paramagnetic metal or doped semiconductor, or a ferromagnetic metal 
above its Curie point,
carrying a charge current in the $x$ direction. The electrons still
carry a spin, and the same scattering mechanism(s) that gave rise to the 
anomalous Hall effect in the magnetic case will scatter electrons
with spin up preferentially in one direction perpendicular to the
flow of current, and spin down electrons preferentially in the
opposite direction. Here we have in mind a slab geometry as usually
used in Hall effect experiments,  and spin up and spin down
directions are defined perpendicular to the plane of the slab. Because
there is equal number of spin up and spin down electrons no charge imbalance
will result, but we argue that a spin imbalance will: there will be 
an excess of up spins on one side of the sample and of down spins
on the opposite side. The situation is depicted schematically in 
Figure 1.

Although it may appear that if there is spin rotational invariance the
spin up and down directions are not well defined, we argue
that the slab geometry naturally defines such directions. The effect can be simply
understood as arising from spin-orbit scattering. Consider\cite{gold} a 'beam'
of unpolarized electrons incident on a spinless scatterer, with potential
\beq
V=V_c(r)+V_s(r) {\vec \sigma}\cdot{\vec L}
\eeq
with ${\vec \sigma}$ and ${\vec L}$ the electron's spin and orbital angular
momentum. The term $V_s(r)$ is the usual spin-orbit scattering potential\cite{gold},
proportional to the gradient of the scattering potential. 
The scattered beam will be spin polarized, with polarization 
vector\cite{gold}
\beq
{\vec P_f}=\frac{f g^*+f^* g}
{|f|^2+|g|^2} {\hat n}
\eeq
where ${\hat n}$ is a unit vector perpendicular
to the scattering plane, in direction $\vec k_i \times \vec k_f$, with
${\vec k_i},{\vec k_f}$ incident and scattered wavevectors. $f$ and $g$ 
are spin-independent and spin-dependent parts of the scattering
amplitude\cite{gold}. ${\hat n}$ has opposite signs for particles
scattered to the right and left of the scatterer, hence there is a left-right
asymmetry to the spin polarization of the scattered beam, whose sign depends
on the sign of $V_s(r)$. In the geometry considered here the
scattering plane is defined by the plane of the slab, since there is
considerably more phase space for scattering in that plane
than perpendicular to it. Furthermore, in a crystal prefered spin directions
may arise from crystalline anisotropy, and it may be useful to consider a single
crystal sample where one such direction is perpendicular to the slab.
Finally, a prefered spin direction is also defined by
the magnetic field generated
by the current flow, which in the slab geometry will point predominantly in the
$+z$ direction on half of the slab along the $y$ direction and in 
the $-z$ direction on the other half. This magnetic field will
contribute an additional spin imbalance, which may add or substract
to the one discussed here depending on the sign of the skew 
scattering mechanism. We will not be interested in this component
of the spin imbalance for reasons discussed below.

In the case of the ordinary Hall effect, the charge imbalance results
in a difference in the Fermi levels of both sides of the sample, and hence
a voltage $V_H$ which can be measured with a voltmeter. In the case
under discussion here, the Fermi levels for each spin electrons will also
be different on both sides of the sample, but the difference will be of
opposite sign for both spins.  How can one detect this spin voltage $V_{SH}$, 
or equivalently the associated spin imbalance?

One possible way would be to measure the difference in magnetization at both edges
of the slab. This may perhaps be achieved by using a superconducting quantum
interference device microscope\cite{blac} with high spatial resolution that can
measure local magnetic fields. However,
it would be necessary to separate the contributions from the effect discussed 
here and the magnetic field generated by the current flow, which is likely
to be difficult because the latter one should be much larger.

A more interesting way follows from the analogy with the ordinary Hall
effect. In that case, if the two edges of the sample are connected by a 
conductor, a charge current will circulate, since the electrons in 
the connecting conductor do not experience the Lorentz force felt by
the electrons in the longitudinal current. Similarly, in our
case we argue that when the edges of the sample are connected a spin
current will circulate. This spin current will be driven solely by
the spin imbalance generated by the skew scattering mechanism(s) affecting
the longitudinal current and not by the component of the spin imbalance
which is due to the magnetic field originating in the current flow.

How does one detect such a spin current? We may use the same principle
that allowed the spin imbalance to be created in the first place.
When the two edges of the sample are connected and a spin current
circulates, a transverse voltage will be generated that can be measured
by a voltmeter. The situation is schematically depicted in Figure 2.

Let us consider some experimental parameters. First, the width of 
the sample $L$ needs to be smaller than the spin diffusion length 
$\delta_s$. $\delta_s$ is the length over which spin coherence is
lost due to scattering processes that do not conserve spin. We will
rely heavily on the seminal work of Johnson and Silsbee\cite{john}
(JS), who studied spin current flow between a ferromagnet and a
paramagnet, aluminum. JS estimated $\delta_s\sim450\mu m$ at
$T=4.3K$ and $\delta_s\sim170\mu m$ at
$T=36.6K$ in their $Al$ sample, which had residual resistivity ratio
(RRR) of about $1000$. We will assume for definiteness
a sample of $Al$ as in the JS experiment, of width $L=100\mu m$, which should
allow for transverse spin coherent transport at least over the
range of temperatures given above. The resistivity of the 
sample for such RRR will be of order 
$\rho=2.7\times10^{-3} \mu \Omega cm$
at low temperatures.

 The 'magnetization' associated with the spin up electrons in the
sample is $M=n_\uparrow \mu_B$,
with $n_\uparrow$ the density of spin up electrons and $\mu_B$ the
Bohr magneton. If only up electrons were present, when a longitudinal current 
density $j_x$ flows an anomalous Hall voltage
\beq
V_H=4\pi R_s L j_x n_\uparrow \mu_B
\eeq
would be generated, with $L$ the width of the sample and $R_s$ the
anomalous Hall coefficient. Equation (4) gives also the 
'spin Hall voltage' for spin up electrons that will be generated,
and an equal one with opposite sign will result for the spin down electrons
in the paramagnetic case. Hence we obtain for the spin Hall voltage
\beq
V_{SH}=2\pi R_s L j_x n \mu_B
\eeq
with $n$ the total conduction electron concentration. 

To obtain an estimate of the magnitude of the effect we 
 will simply assume that $R_s$ is
the same as the free electron ordinary Hall coefficient of 
$Al$, $R_o=-1/nec=-3.45\times10^{-11}m^3/C$ .
As mentioned above, values of the anomalous Hall coefficient
tend to be larger than those of the ordinary one. For a current
density $j_x=6\times 10^6 A/m^2$ as used in the JS experiment
 Eq. (5) yields a spin Hall voltage $V_{SH}=22nV$. 

When we connect the two edges of the sample by a transverse metal
strip, a spin current will
flow in that strip. Assuming that the resistivity for the spin current is the
same as that for the charge current we have for the current for
each spin
\beq
j_\sigma=\frac{V_{SH}}{\rho L}
\eeq
which yields for the parameters under consideration
$j_\sigma=8.0\times 10^6 A/m^2$. Assuming the same skew scattering 
mechanism operating
on the transverse sample, the resulting spin Hall voltage due to
this spin current is
\beq
V_{SH}^\sigma=4\pi R_s l j_\sigma n_\sigma \mu _B
\eeq
with $l$ the width of the transverse strip.
Now however because spin up and down currents circulate in 
opposite directions the spin voltages add, giving rise to a
real voltage due to the spin current
$V_{SC}=2V_{SC}^\sigma$,
that can be detected by an ordinary voltmeter. The voltage due to the
spin current is then, from Eqs. (5), (6) and (7)
\beq
V_{SC}=8\pi^2 R_s^2l \frac{(n\mu_B)^2}{\rho} j_x.
\eeq 
Note that the transverse width $L$ has dropped out in Eq. (8), because
even though it gives larger spin voltage $V_{SH}$ it also increases
the resistance to the spin current in the transverse direction.
Still, a dependence on $L$ is implicit in Eq. (8) since when $L$
becomes comparable to or larger than the spin diffusion length
$\delta_s$, $V_{SC}$ will decrease. Neither does
the thickness of the transverse layer enter in Eq. (8): a thicker layer would 
increase the spin current but not the current density. For the
parameters under consideration here, assuming for example
$l=100\mu m$ , Eq. (8) yields $V_{SC}=58 nV$, easily measurable.
 In the more general
case where the transverse strip is of different composition and/or
purity than the longitudinal strip Eq. (8) becomes
\beq
V_{SC}=8\pi^2R_{s1}R_{s2} l \frac{n_1n_2 \mu_B^2}{\rho_2}j_x
\eeq
where indices $1$ and $2$ refer to longitudinal and transverse
strips respectively.

Figure 3 shows top and side views of the sample envisaged. A thin
insulating layer should be deposited on top of the sample
(longitudinal strip) of width $L$, and small contact areas should be etched
to expose the sample surface and allow for metallic contact
 between the longitudinal and transverse
strips. Then, a thin transverse strip of width 
$l$ should be deposited on the insulator such that it also covers the contact
areas. The length of the
contacts along the $x$ direction should be sufficiently small that no
significant voltage drop should occur on them
due to the longitudinal current, which would be transmitted to the
transverse strip. The voltage drop along a contact of length
$l_c$,
\beq
V_d=l_c\rho j_x,
\eeq
should be substantially smaller than the signal $V_{SC}$. For the parameters
used as example here, $V_d=0.2nV$  for a contact width
$l_c=1\mu m$. Also, a spurious voltage may arise if the two contacts are not
perfectly alligned. Again, if the contacts are offset by $\Delta x$ the magnitude of the
spurious voltage will be at most Eq. (10) with $\Delta x $ replacing $l_c$,
so for our parameters $V_d\sim 1nV$ if $\Delta_x=5\mu m$. Note also that
a smaller resistivity $\rho$ both increases the signal voltage Eq. (8)
and decreases the spurious voltage Eq. (10).
Finally, the resistance of the contacts should be much smaller than the resistance
of the transverse strip in order for Eq. (8) to remain valid.
This argues for a thin transverse strip (large resistance) and
a thin insulating layer (smaller contact resistance along
width of layer). It would appear to be simple to achieve a contact
resistance at least two orders of magnitude smaller than the transverse strip
resistance.

Note also that the sign of the expected signal $V_{SC}$, as indicated in 
Fig. 2, is opposite to the voltage $V_d$ that would arise from voltage
drop across the contacts. As long as the
signs of the anomalous Hall coefficients $R_{s1}$ and $R_{s2}$ are the
same, so in particular for $R_{s1}=R_{s2}$, the sign of the spin current 
voltage $V_{SC}$ will always be as indicated in Fig. 2, that is, 
$V_{SC}$ drives a current in direction opposite to the primary current
$j_x$.
Thus a measurement of $V_{SC}$ for the case where the
longitudinal and transverse strips are of the same material provides
no information on the sign of $R_s$. If the signs of
$R_{s1}$ and $R_{s2}$ are opposite however the sign of the voltage
$V_{SC}$ would be reversed. 

Application of a magnetic field in a direction parallel
to the plane of the strips will lead to precession of the spins and destruction of 
the spin 
polarization for a characteristic field $(\gamma T_2)^{-1}$, 
as discussed by Johnson and Silsbee\cite{john}, with
$\gamma$ the gyromagnetic ratio and
$T_2$ the spin relaxation time of the conduction electrons.
Thus it would lead to suppression of the spin current voltage.
The sensitivity of the signal to an applied magnetic field in the plane
of the strips would provide  direct evidence for the role of electron spin.
For the case of the $Al$ sample of JS the signal would be entirely
suppressed for magnetic fields in the range of $20$ to $50$ Gauss
depending on temperature\cite{john}.

The sign of the expected effect as indicated in Figure 2
is however the same as would be obtained from a 'drag effect' of the current in
the lower strip on the upper strip. Such effects, which
may arise from electron-phonon\cite{shoc} or electron-electron\cite{rojo}
 interactions,
have been seen in doped semiconductor structures\cite{drag1,drag2}
, and they could contribute
to the effect discussed here. However, the drag effect does not require
contact between the lower and upper strips, should not vary with applied
magnetic field in the plane, and should sensitively depend on the thickness of 
the insulating layer (in our case sensitivity to insulating layer thickness 
might only enter insofar as it could affect the contact resistance
between lower and upper strips). These differences should make it 
possible to differentiate one effect from the other. Then it is also
possible that the drag effect that has been observed\cite{drag1,drag2}
 occurred in the
presence of some contact between the two layers or of tunneling
 that allowed spin current
to flow and thus had a contribution from the effect discussed here. To our
knowledge sensitivity to applied magnetic field in the plane 
was not checked in those cases.

In the presence of a magnetic field $B$ in the $z$ direction an ordinary
Hall voltage across the longitudinal strip will be generated, which will
cause charge current to flow across the transverse strip and give another 
contribution to the voltage generated by the spin current Eq. (8). The 
total voltage across the transverse strip will be
\beq
V_t(B)=(R_o^2 B^2+R_s^2 B_{eq}^2)\frac{l}{\rho}j_x ,
\eeq
with
$B_{eq}=4\pi n\mu_B /\sqrt{2}$.
The sign of the ordinary contribution to the voltage across the transverse strip
is the same as that of the
spin Hall effect, independent of the sign of the applied magnetic
field $B$ and of the sign of $R_o$. Even if experimental resolution
impedes accurate measurement of $V_t$ for $B=0$, it may be
possible to extract the effect discussed here from extrapolation
of results for $V_t(B)$ to $B=0$.

In conclusion, the experiment proposed here, if successful, would
achieve the following: (1) It would provide a 
realization of spin current flow in the absence of charge current
flow; (2) it would demonstrate that flow of a spin current results in the
generation of a transverse electric field;\cite{hirs2}
(3) it would show the generation of spin imbalance in
a paramagnetic metal when a charge current circulates; 
(4) it would establish the existence of a skew scattering
mechanism in a paramagnetic metal; (5) it would
provide information on the magnitude and temperature dependence
of the anomalous Hall coefficient $R_s$ in a paramagnetic
metal; (6) measurement of the dependence of the voltage
$V_{SC}$ on strip width, temperature and magnetic field
would provide information on processes that lead to loss of spin
coherence; (7) measurement of dependence of the magnitude of
$R_s$ on sample purity and temperature would provide information
on the scattering mechanism(s) responsible for $R_s$, and in
particular on whether a periodic potential by itself can give rise
to an anomalous Hall effect; (8) assuming a 
known sign for $R_s$ of the longitudinal strip, it would allow determination
of the sign of $R_s$ of the transverse strip. These and other
findings resulting from this experiment could have practical 
applications in the field of spin electronics\cite{prin}. 
Even though we
discussed the effect here assuming a metallic sample it is possible that
semiconducting samples
may allow for easier detection of this effect. Furthermore, it would be
of interest to study this effect in the limit where the strips in figure 3 
are two-dimensional, as in the electron bilayer systems in GaAs double
quantum well structures extensively used in studies of the quantum
Hall effect.\cite{eise}

\begin{figure}
\caption { The charge carriers are assumed to be electron-like.
In the Hall effect, the Lorentz force on the moving charges causes charge 
imbalance, in the spin Hall effect skew scattering of the moving
magnetic moments causes spin imbalance, in direction perpendicular
to the current flow.
In the Hall effect the  Fermi levels for up and down electrons
are the same, and the difference in the Fermi levels at both edges of the
sample is the Hall voltage $V_H$. In the spin Hall effect the difference
in the Fermi levels for each spin at both edges of the sample is
$V_{SH}$ but it is of opposite sign for spin
up and down electrons.}
\label{Fig. 1}
\end{figure}

\begin{figure}
\caption {  A transverse strip of width $l$ connects both edges
of the slab. A spin current will flow and skew scattering will
cause negative charge to accumulate on the left edge (upstream from 
the primary current $j_x$). A charge imbalance will result and
an electric potential that can be measured with a voltmeter.}
\label{Fig. 2}
\end{figure}

\begin{figure}
\caption {  Top view (along the $-z$ direction ) and side view
(along the $+y$ direction) of the sample envisaged for detection of
the spin Hall effect. The voltage $V$ measured by the voltmeter will
be the spin current voltage Eq. (6) in the absence of applied magnetic field
or the voltage $V_t(B)$, Eq. (9), in the presence of a magnetic field
$B$ in the $z$ direction.
}
\label{Fig. 3}
\end{figure}


\begin{references}
\bibitem{hirs} In a separate paper we will argue that in fact an
anomalous Hall effect should exist even in a perfectly periodic lattice
at zero temperature.
\bibitem{hirs2} Another situation where these phenomena are predicted to occur 
was discussed by the present author  in: Phys. Rev. B {\bf 41}, 6828 (1990); 
Phys. Rev. B {\bf 42}, 4774 (1990).
\end{references}
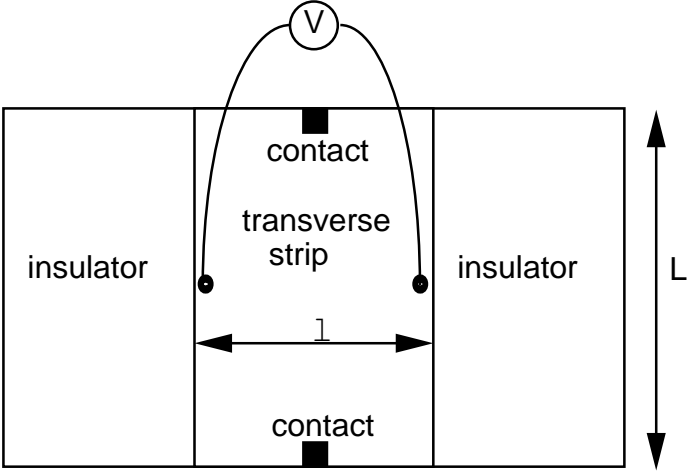
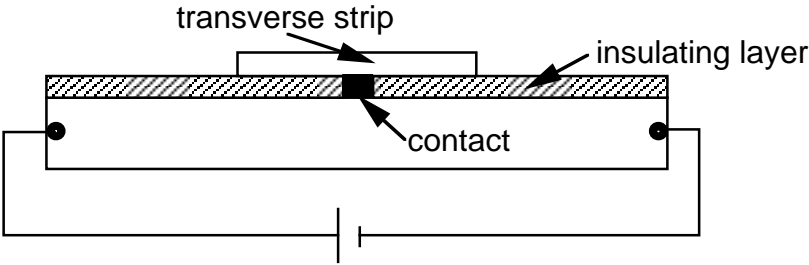
\end{document}